\documentclass{aastex}
\usepackage{natbib,graphicx,latexsym}
\usepackage{emulateapj5}

\newcommand{\HST}{{\sl HST}}

\newcommand{\perone}{\mbox{$^{-1}$}}

\newcommand{\etal}{et al.}

\newcommand{\ie}{i.e.}

\newcommand{\kms}{\hbox{km~s$^{-1}$}}

\newcommand{\Ho}{\mbox{$H_0$}}

\newcommand{\Ks}{\mbox{$K_S$}}
\newcommand{\Kp}{\mbox{$K^{\prime}$}}

\newcommand{\VmI}{\mbox{$V\!-\!I_c$}}

\newcommand{\mbarK}{\mbox{$\overline{m}_K$}}
\newcommand{\MbarK}{\mbox{$\overline{M}_K$}}

\newcommand{\varsky}{\mbox{$\sigma_{sky}^2$}}
\newcommand{\varglob}{\mbox{$\sigma_{GC}^2$}}

\newcommand{\bk}{\mbox{${\bf k}$}}

\slugcomment{ApJ Letters, in press (June 2001)}

\shorttitle{Liu \& Graham}
\shortauthors{IR SBF Distance to the Coma Cluster}

\begin{document}

\title{Infrared Surface Brightness Fluctuations of the 
Coma Elliptical NGC~4874\\
and the Value of the Hubble Constant\altaffilmark{1}}

\author{\sc Michael C. Liu\altaffilmark{2}  and James R. Graham}
\affil{Astronomy Department, University of California, Berkeley, CA
94720} 

\altaffiltext{1}{Based in part on observations obtained at the W.M. Keck
Observatory, which is operated as a scientific partnership among the
California Institute of Technology, the University of California, and the
National Aeronautics and Space Administration.}

\altaffiltext{2}{Currently Beatrice Watson Parrent Fellow, Institute for
Astronomy, University of Hawai`i, 2680 Woodlawn Drive, Honolulu, HI
96822.}

\email{mliu@ifa.hawaii.edu}

\begin{abstract}

We have used the Keck~I Telescope to measure $K$-band surface brightness
fluctuations (SBFs) of NGC~4874, the dominant elliptical galaxy in the
Coma cluster.  We use deep \HST\ WFPC2 optical imaging to account for
the contamination due to faint globular clusters and improved analysis
techniques to derive measurements of the SBF apparent magnitude.  Using
a new SBF calibration which accounts for the dependence of $K$-band SBFs
on the integrated color of the stellar population, we measure a distance
modulus of $34.99\pm0.21$~mag ($100 \pm 10$~Mpc) for the Coma cluster.
The resulting value of the Hubble constant is $71 \pm
8$~\kms~Mpc\perone, not including any systematic error in the \HST\
Cepheid distance scale.

\end{abstract}



\section{Introduction}

The Coma cluster (Abell 1656) is an important rung in the cosmic
distance ladder.  
It is the nearest of the very rich Abell clusters (richness class of 2
[\citealp{1958ApJS....3..211A}]) and thus contains many luminous
galaxies which can be targeted for distance measurements.  Also, the
cluster redshift ($\approx$7000~\kms) is large enough that its
cosmological recession is not significantly perturbed by smaller-scale
peculiar velocities \citep[e.g.][]{1995PhR...261..271S}.  For both these
reasons, several methods have been used to measure the distance to Coma
and thereby determine the Hubble constant (\Ho).  These include Type~Ia
supernovae \citep{1990ApJ...350..110C}, the Tully-Fisher relation
\citep{1998AJ....116.2632G}, the Fundamental Plane
\citep{1995MNRAS.276.1341J, 1997MNRAS.289..847D}, the globular cluster
luminosity function \citep[e.g.,][]{2000ApJ...533..125K}, and surface
brightness fluctuations \citep{1997ApJ...483L..37T, jen01}.

Surface brightness fluctuations (SBFs) are an appealing distance
indicator because they have a well-understood physical basis --- they
arise from Poisson statistical fluctuations in the number of stars in a
resolution element (\ie, the seeing disk).
Over the past decade, optical SBFs have been used to measure distances
to elliptical galaxies and the bulges of spiral galaxies
\citep{1992PASP..104..599J, 1999phcc.conf..181B}. $I$-band (0.8~\micron)
SBFs vary between galaxies by up to one magnitude, but the variations
are well-correlated with \VmI\ color so they can be compensated.  More
recently, near-infrared (IR) SBFs have been shown to be useful distance
indicators.  Because cool giant stars dominate the spectral energy
distributions of ellipticals, SBFs are brighter in the near-IR and hence
can be detected at greater distances.  Unresolved globular clusters are
the dominant contaminant for SBF measurements at large distances, but
since the clusters are bluer than the old stars in ellipticals, the
contamination is much smaller at IR wavelengths.  $K$-band (2.2~\micron)
SBFs have been measured for galaxies in several clusters
\citep{1993ApJ...410...81L, 1994ApJ...433..567P, 1998ApJ...505..111J,
1999ApJ...510...71J, mei2001, liu01}.

The core of the Coma cluster is dominated in luminosity by two
supergiants, the elliptical NGC 4889 ($B_T=12.53$~mag) and the cD NGC
4874 ($B_T=12.63$~mag; \citealp{RC3}).  Like many Abell clusters, Coma
possesses significant substructure \citep[e.g.,][]{1993AJ....105..409D,
1993ApJ...413..492M}. However, NGC~4874 appears to be special for two
reasons: it is located at peak of the diffuse X-ray emission
\citep{1993MNRAS.261L...8W}, and it has a strong nuclear radio source
characteristic of many central giant ellipticals
\citep{1987ApJ...315L..29H}.  Therefore, it is considered to reside at
the heart of the gravitational potential of the cluster
\citep{1990AN....311...89B,1996ApJ...458..435C}.

In this paper, we measure the distance to NGC~4874 using $K$-band
SBFs. We have recently developed new theoretical models for SBF studies
\citep[][hereinafter Paper~I]{bc2000sbf}. Also, we have completed a new
calibration of $K$-band SBFs using a large sample of early-type galaxies
in nearby clusters \citep[][hereinafter Paper~II]{liu01}.
We have found that $K$-band SBFs vary considerably between galaxies but,
like $I$-band SBFs, do so in a predictable fashion as they are
correlated with \VmI\ galaxy color.  In this paper, we use these results
with new SBF data from the Keck~I Telescope to derive an accurate distance
to NGC~4874 and consequently a measurement of \Ho.


\section{Observations} \label{sec5:obs}


We observed NGC~4874 at the 10-meter Keck~I telescope using the facility
instrument NIRC \citep{nirc} with the standard $K$-band filter
(2.0--2.4~\micron).  The camera uses a Santa Barbara Research
Corporation 256 $\times$ 256 InSb array and has a
$38\arcsec\times38\arcsec$ field.  Table~\ref{tbl:comasbf} summarizes
the observations.  Both runs were photometric, with 0\farcs5 FWHM
seeing.  Images on the galaxy were interlaced with blank sky fields and
taken in an ABBA pattern.

We obtained wider-field \Kp-band (1.9--2.3~\micron) images on 28 June
1995 UT using the facility near-IR camera LIRC2
\citep{1995IAUS..167...79G} on the Shane 3-m telescope at Lick
Observatory.  LIRC2 employs a Rockwell International $256\times256$
HgCdTe NICMOS-3 array and has a plate scale of 0\farcs38~pixel\perone.
Data were obtained in a similar fashion as the Keck data, taking
interlaced pairs of sky and galaxy images.  Conditions were photometric
with a typical seeing FWHM of 1\farcs1. For all the runs we observed the
faint IR standards of \citet{cas92} as flux calibrators. Hence, our
resulting magnitudes are Vega-based.


Details of the data reduction appear in Paper~II.
We subtracted an average bias from the images. Using twilight sky
images, we constructed flat fields with an iterative-fitting algorithm
which separated the flat field from the non-uniform thermal emission on
the array.  A preliminary sky subtraction was performed to identify
astronomical objects.  Then for each image, we made a running sky frame
from the prior and subsequent images of blank sky, excluding any
astronomical objects.  The sky brightness changes temporally so we
scaled the running sky frames to the median counts in unsubtracted
galaxy images, excluding a circular region centered on the galaxy.  We
used the same masks for scaling the subtraction of the blank sky
images. This ensured the blank sky images and galaxy images were reduced
identically.
NIRC images of bright sources suffer from ``bleeding'': they have a
positive horizontal trail which exponentially weakens along the readout
sequence.
We used images of bright stars to model the bleeding and remove it.
We also used a software mask to exclude a faint in-focus ghost
of the exit pupil (the secondary mirror) present in all the images.
We used the galaxy to register the individual frames and averaged to
assemble a final mosaic.

%
Finally, we used the Lick images to determine the remaining DC sky
background in the Keck images.  Being a cD-type galaxy, NGC~4874 has a
rather shallow light profile; hence the common technique of assuming the
galaxy follows a de~Vaucouleurs $r^{1/4}$ profile is ill-suited.  We
extracted azimuthally averaged profiles from the Lick and Keck images,
and fitted for the DC level in the Keck images.  To estimate the offset
between the slightly different Lick and Keck filters, we used spectra of
solar-metallicity M0--M5 giants from \citet{1998PASP..110..863P}, which
should be well-suited for this purpose \citep{1978ApJ...220...75F}.  Our
synthesized $(\Kp\!-\!K)$ color was $-0.015\pm0.003$~mag, including
accounting for the filters' $k$-corrections at the redshift of
Coma.\footnote{Our synthesized color is notably different than what
would result from the transformation given in
\citet{1992AJ....103..332W}, which is $\Kp-K = (0.18\pm0.04)(H-K)$.
However, as they point out, their transformation was derived from a
sample of A~stars and M~dwarfs.  Neither of these have the strong
2.3~\micron\ CO absorption band seen in giant stars, which dominate the
near-IR light of elliptical galaxies.  The CO feature is within the
standard $K$-band filter, but not in the \Kp-band filter; hence, the
sign of our synthesized correction is as expected.}
We checked our final calibrated images against the photometry of
\citet{1979ApJS...39...61P} taken in a 14\farcs9 diameter aperture. The
agreement was excellent, with our photometry being on average
$0.014\pm0.008$~mag brighter.


\section{SBF Measurements} \label{sec5:measurements}



Our SBF measurement methods are similar to
\citet{1990AJ....100.1416T}. A complete discussion appears in Paper~II.


The mean surface brightness of NGC~4874 was modeled by fitting for the
harmonic content of isophotes.
Globular clusters and background galaxies fainter than the detection
limit are unresolved point sources in our images. Therefore, they
contribute to the fluctuations of the galaxy surface brightness.  The
effect is substantial, because nearly the entire globular cluster
population of the galaxy is unresolved in the $K$-band images (see
below).  In order to avoid this contamination, we identified the
positions of the globular clusters in our $K$-band data using archival
deep \HST\ $V$-band imaging of NGC~4874 \citep{2000ApJ...533..125K},
which reach down to the peak ($V\approx27.9$~mag) of the globular
cluster luminosity function (GCLF).

We then constructed a software mask defining an annular region and
excluding the globular clusters.  For the 1998 Keck data, we used two
annuli, 3--9\arcsec\ and 9--12\arcsec, chosen to have comparable area.
The 1995 Keck data are of lower quality: the total integration was
shorter, and at the time the instrument had much higher noise.
Therefore, we used only the 3--9\arcsec\ annulus for this data.  The
model-subtracted galaxy image was multiplied by this mask, and the
Fourier power spectrum of the central 256$\times256$ pixel region
determined.

\begin{figure*}[t]
\vskip -1in
\centerline{\includegraphics[width=4in,angle=90]{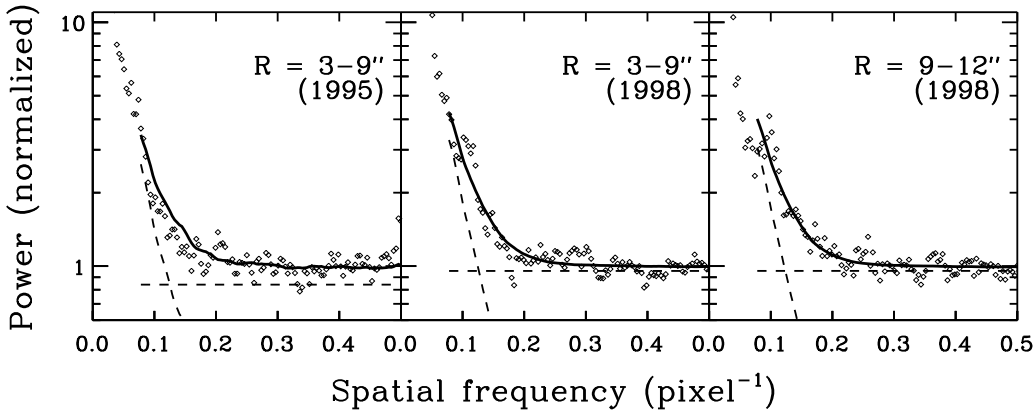}}
\vskip -0.7in
\caption[$K$-band fluctuation power spectra for NGC~4874]{\small
$K$-band fluctuation power spectra for NGC~4874.  The galaxy power
spectra were fitted by the sum (solid line) of a scaled version of the
PSF ($P_0 \times E(k)$) and a constant ($P_1$); the dashed lines show
the contributions from these two components.  We did the fit with the
two-dimensional power spectra, using spatial frequencies of
0.08~pixel\perone\ to the Nyquist frequency. One-dimensional azimuthal
averages are plotted to represent the results, with the measurement
annuli and year of observation labeled. \label{fig:pspec-n4874}}
\vskip -0.2in
\end{figure*}

The power spectrum of the model-subtracted galaxy image has two
components: (1) rising power at low wavenumbers due to the fluctuations
convolved by the point spread function (PSF), and (2) white noise due to
the Poisson shot noise.  The power spectrum $P(\bk)$ can be represented
as
\begin{equation}
P(\bk) = P_0 \times E(\bk) + P_1,
\label{eqn5:pspec}
\end{equation}
where $P_0$ is the total variance per pixel, $P_1$ is white noise, and
$E(\bk)$ is the expectation power spectrum.  $E(\bk)$ is the convolution
of the PSF power spectrum with the power spectrum of the software mask
times the square-root of the galaxy model.  It accounts for the PSF, the
radial variation in the SBFs, and the effect of the software mask on
$P(\bk)$ (see \citealp{liuthesis}).
Using $E(\bk)$, we fitted the two-dimensional power spectra to solve for
$P_0$ and $P_1$ (Fig.~\ref{fig:pspec-n4874}).

For the very lowest wavenumbers, flat-fielding and sky-subtraction
errors produce extra power which contaminates the SBF signal.  We
examined the power spectra of the blank sky fields, which were observed
and reduced in an identical fashion to the galaxy images. We found that
wavenumbers of \mbox{$k\lesssim20$} (0.08~pixel\perone), corresponding
to spatial scales of $\gtrsim4\times$~FWHM, had significant rising
power.  Therefore, we used only $k=20$ to $k=128$, the Nyquist
frequency, when fitting the galaxy power spectra.

We analyzed the power spectrum of each quadrant of the galaxy image
independently. We used the average of these four fits to obtain $P_0$
and computed the standard error as the uncertainty in $P_0$.
In Paper~II, we ran Monte Carlo tests to verify that these
quadrant-derived errors are accurate. Each Keck data set had only one
suitable PSF star. Thus we adopted a PSF mismatch error of 0.08~mag,
which we determined in Paper~II from our Fornax cluster sample.


$P_0$ is the sum of the variance from the galaxy's stars, which is the
desired signal, along with the variance from unresolved astronomical
sources and instrumental effects.  We quantified the effect of
unresolved globular clusters using an analytic representation for the
GCLF, namely a Gaussian with three parameters (width, peak magnitude,
and normalization) as measured by \citet{2000ApJ...533..125K} using the
\HST\ $V$-band data.  To convert to $K$-band, we adopted a color of
$V\!-\!K=2.28$, the average of the M31 and Milky Way globular clusters
\citep{2000AJ....119..727B}.  (The optical colors of the NGC~4874
clusters are similar to the Milky Way population, being relatively blue
and having a moderately narrow dispersion
[\citealp{2000ApJ...533..137H}].)
We then computed the variance per pixel due to the globular clusters
below the detection limit of the \HST\ optical images, \varglob, using
eqn.~5 of \citet{1995ApJ...442..579B}.
Note that without the \HST\ data, the variance from the undetected
globular clusters would be comparable to that from the stellar SBFs, the
signal we seek.

Instrumental signatures from flat-fielding and sky subtraction errors
add variance to the lowest wavenumbers.  By restricting our fits to
$k\geq20$, we reduced the systematic errors.  However, these effects
might still contaminate the fitting, since they are not step functions
in the power spectrum.  This is particularly true for NIRC, which has
spatially complex flat fields and bias frames. To quantify the amount of
leakage into the wavenumbers used for fitting, we used the same
techniques on the blank sky fields and measured the blank sky variance
per pixel \varsky\ at numerous locations on the images.  We used the
scatter in the results to find the error in \varsky, around 10--15\%.
While most of \varsky\ arises from instrumental signatures, some comes
from real astronomical sources, namely background galaxies fainter than
the detection limit.  Therefore, unlike the usual SBF analysis, there
was no need to compute the residual variance due to unresolved galaxies
as this was included in \varsky. (Using the counts of
\citealp{1998ApJ...505...50B}, the variance from the galaxies is
predicted to amount to $\approx$10\% of the total variance.)

Finally, the total amplitude of the power spectrum $P_0$ was corrected
by subtracting the contaminating variance $P_r$.  The quantity $P_r$ is
the sum of $P_{GC}$ and $P_{sky}$, which are \varglob\ and \varsky\
divided by the mean galaxy surface brightness per pixel in the
measurement region, respectively. The exceptional depth of the \HST\
optical data meant that residual variance from the unmasked globular
clusters was negligible ($\approx$0.002~mag).  The blank sky variance
was substantial, about half of the total variance measured in the galaxy
images. However, the measurements appear to be robust.  For the 1998
Keck data, \varsky\ measured in the two different annuli showed
excellent agreement.  In the 1995 Keck data, \varsky\ was substantially
larger as expected by the higher noise levels then. After subtracting
this variance, the resulting $K$-band SBF apparent magnitude (\mbarK)
agrees well with those from the 1998 data.

The remaining quantity ($P_0\!-\!P_r$) is the variance due solely to the
stellar SBFs.  To convert to $\mbarK$, we used the measured photometric
zero point and applied two corrections.  We corrected for 0.003~mag of
extinction \citep{1998ApJ...500..525S}.  Using the solar-metallicity
models of Paper~I, we also applied very small $k$-corrections to \mbarK\
and \VmI\ to account for the redshifting of the galaxy light.
The final errors on \mbarK\ comprise the quadrature sum of the errors in
the photometric calibration (0.02~mag), PSF uncertainty (0.08~mag), and
measurement errors in the variances (\varglob, \varsky, and $P_0$).
Table~\ref{tbl:comasbf} presents our results.


\section{Results} \label{sec5:results}


In order to compute the distance modulus, we use our $K$-band SBF
calibration from Paper~II.  This calibration uses 24 early-type galaxies
in nearby clusters with high-quality SBF data; the zero point is based
on $I$-band SBF distances to the bulges of six nearby luminous spiral
galaxies \citep{sbf4} which have \HST-derived Cepheid distances from
\citet{2000ApJS..128..431F}.
(The effect of using the new Cepheid distances by \citealp{fre01} is
$\sim$0.05~mag --- see Paper~II.) The calibration is for the \Ks-band
(2.0--2.3~\micron; \citealp{1995ApJS...96..117M}) filter, which has a
slightly bluer bandpass than the $K$-band filter used here.  To estimate
the effect of this difference, we use the theoretical models of Paper~I
with ages of 3--12~Gyr and metallicities of solar and slightly sub-solar
($Z=0.02$ and 0.008); these span the SBF observations to date (Papers~I
and II).  The mean $(\Ks\!-\!K)$ offset is effectively zero (0.004~mag)
with an rms of 0.02~mag.  Thus, we assume no difference but include the
rms as an additional source of error. The resulting calibration of the
$K$-band SBF absolute magnitude is:
\begin{equation}
\MbarK = (-5.84\pm0.04) + (3.6\pm0.8) [(\VmI)_0 - 1.15]\ .
\end{equation}
This gives us the expected \MbarK\ for NGC~4874 given its
($k$-corrected) \VmI\ color.  A weighted average of the distance moduli
from the three data sets is $34.99\pm0.15$~mag.  However, the intrinsic
(cosmic) scatter about the mean relation is non-zero; in Paper~II, we
conservatively estimated it to be 0.15~mag.
We add this in quadrature to determine a final distance modulus of
$34.99\pm0.21$~mag ($100\pm10$~Mpc).

Our errors do not include any systematic errors in the \HST\ Cepheid
distance scale. There are several potential sources of error including
the assumed distance to the Large Magellanic Cloud, the metallicity
dependence of the Cepheid period-luminosity relation, and the
photometric calibration of the \HST\ WFPC2 instrument. Any changes in
the \HST\ Cepheid scale will affect our results by changing the $K$-band
SBF calibration.

Our result is comparable to previous $K$-band SBF determinations of the
distance to Coma.  \citet{1997ApJ...483L..37T} used $I$-band
observations from \HST\ to derive a distance of $102\pm14$~Mpc to the
Coma elliptical NGC~4881. \citet{1999ApJ...510...71J} measured a
$K$-band SBF distance of $85\pm10$~Mpc to the other Coma supergiant
galaxy NGC~4889.  Both of these previous results were based on
observations with much lower S/N, $(P_0\!-\!P_r) < 0.7 P_1$, at least a
factor of 10 worse than our measurements here.


To determine \Ho, we adopt a mean velocity for the Coma cluster in the
reference frame of the cosmic microwave background of $7186\pm428$~\kms\
from \citet{1992ApJ...396..453H}.  At small redshifts, the \Ho\ is
related to the distance modulus by $(m-M) = 25 + 5\log(cz/H_0) +
1.086(1-q_0)z$. The last term accounts for cosmological curvature, and
its effect on \Ho\ is negligible: $\leq2\%$ at the redshift of Coma for
$q_0$ ranging from --1 to +1.  Our resulting Hubble constant is
\hbox{$71\pm8$~\kms~Mpc\perone}.




\acknowledgments

Keck Observatory was made possible by the generous support of the W.\
M.\ Keck Foundation.  We thank the staffs of Lick and Keck Observatories
for their help, especially Chuck Sorenson, Wendy Harrison, and Wayne
Earthman.  We also thank Bill Harris and J.J. Kavelaars for making their
NGC~4874 results available.
This research was supported by NSF grant no.\ AST-9617173 and \HST\ NASA
grant no. GO-07458.01-96A to the authors.  M.~Liu is also grateful for
support from the Beatrice Watson Parrent Fellowship at the University of
Hawai`i.

\clearpage





\clearpage



%


\begin{deluxetable}{lccccccccccc}
\tablecaption{NGC 4874 $K$-band SBF Measurements \label{tbl:comasbf}}
\tablewidth{0pt}
\tabletypesize{\small} 
\rotate

\tablehead{
\colhead{Date}            & 
\colhead{$t_{int}$}       & 
\colhead{Radius}  & 
\colhead{$\langle\mu_K\rangle$\tablenotemark{a}} & 
\colhead{\VmI\tablenotemark{a}}          & 
\colhead{$P_{GC}/P_0$\tablenotemark{b}}      &
\colhead{$P_{sky}/P_0$\tablenotemark{b}}       &
\colhead{$(P_0\!-\!P_r)/P_1$\tablenotemark{c}} &
\colhead{$\mbarK$\tablenotemark{d}} \\
\colhead{}  & 
\colhead{(s)}  & 
\colhead{(\arcsec)}  & 
\colhead{(mag/$\Box$\arcsec)} & 
\colhead{(mag)}  & 
\colhead{}  & 
\colhead{}  & 
\colhead{}  & 
\colhead{(mag)}
}

\startdata


18 Mar 1995 & 1920 & 3--9  & 16.68 &  1.219 $\pm$ 0.015  &  0.00 $\pm$ 0.13  &  0.69 $\pm$ 0.17  &  6.7 $\pm$ 1.4  & 29.52 $\pm$ 0.32 \\
															      
19 Mar 1998 & 3600 & 3--9  & 16.68 &  1.219 $\pm$ 0.015  &  0.00 $\pm$ 0.14  &  0.48 $\pm$ 0.20  &  9.3 $\pm$ 2.2  & 29.47 $\pm$ 0.22 \\
19 Mar 1998 & 3600 & 9--12 & 17.49 &  1.198 $\pm$ 0.015  &  0.00 $\pm$ 0.10  &  0.58 $\pm$ 0.16  &  7.2 $\pm$ 1.3  & 29.18 $\pm$ 0.24 \\

%

\enddata


\tablenotetext{a}{Mean $K$-band surface brightness and \VmI\ color of
the region used for SBF measurement.  The \VmI\ data are from Harris
\etal\ (2000) and have had $k$-corrections applied.}

\tablenotetext{b}{The fractional contamination to the total SBF signal
$P_0$. $P_{GC}$ is the variance from globular clusters which are
undetected in the \HST\ optical imaging. $P_{sky}$ is the variance
measured from images of blank sky, which arises from instrumental noise
and unresolved background galaxies.  Both $P_{GC}$ and $P_{sky}$ are
normalized to the mean surface brightness in the measurement region.
The quoted errors include the error in $P_0$, which dominates the error
in $P_{GC}/P_0$.}

\tablenotetext{c}{The signal-to-noise of the stellar SBF signal.  $P_r$
is the total residual variance and is the sum of $P_{GC}$ and $P_{sky}$.
The difference $(P_0-P_r)$ is the stellar SBF variance.  $P_1$ measures
photon shot noise, the flat component of the power spectra plots.}

\tablenotetext{d}{The final SBF apparent magnitudes. The values have
been corrected for extinction and have had $k$-corrections applied.}


\end{deluxetable}

 
\end{document}